\documentclass[fleqn,usenatbib]{mnras}
\usepackage[T1]{fontenc}
\usepackage{ae,aecompl}

\usepackage{graphicx}
\usepackage{amsmath}
\usepackage{amssymb}

\title[Cross-correlation analysis of CMB with foregrounds]{Cross-correlation analysis of CMB with foregrounds for residuals}

\author[Pavan K. Aluri and Pranati K. Rath]{
Pavan K. Aluri$^{1}$\thanks{E-mail: aluri@iucaa.in}
and Pranati K. Rath,$^{2}$\thanks{E-mail: pranati@iopb.res.in}
\\
$^{1}$Inter-University Centre for Astronomy and Astrophysics, Pune-411007, India\\
$^{2}$Institute Of Physics, Bhubaneswar-751005, India\\
}

\date{Accepted XXX. Received YYY; in original form ZZZ}

\pubyear{2016}

\begin{document}
\label{firstpage}
\pagerange{\pageref{firstpage}--\pageref{lastpage}}
\maketitle

\begin{abstract}
In this paper, we try to probe whether a clean CMB map obtained
from the raw satellite data using a cleaning procedure is sufficiently
clean. Specifically we study if there are any foreground residuals still
present in the cleaned data using a cross-correlation statistic.
Residual contamination is expected to be present, primarily, in the galactic plane
due to the high emission from our own galaxy. A foreground mask is
applied conventionally to avoid biases in the estimated quantities of
interest due to foreground leakage. Here, we map foreground
residuals, if present, in the unmasked region i.e., outside a CMB analysis mask. Further
locally extended foreground-contaminated regions, found eventually,
are studied to understand them better. The few contaminated regions thus identified
may be used to slightly extend the available masks to make them more stringent.
\end{abstract}


\section{Introduction}
With the unprecedented measurements of the cosmic microwave background (CMB)
temperature anisotropies, cosmology has plunged into a precision era \citep{wmap1yr,planck13}.
This high resolution CMB data has been in turn used to probe some of the underlying
assumptions of the standard cosmological model such as isotropy and Gaussianity.
Since the release of CMB data from WMAP's first year observations \citep{wmap1yr}, many
anomalous deviations from the standard model expectations have been reported
(see \cite{w7yranom,w9yrmaps,plk13anom,plk15anom} and the references there in).
Some of these anomalies were studied in the context of foreground residuals that may
still be present in the \emph{clean} CMB maps, for example, in
\cite{slosar04,abramo06,bernui06,costa06,rakic06,copi07,babich08,gruppuso09,cabella10,
short10,aluri11,cruz11,aluri12,lacasa12,chingangbam13,kovacs13,liu14,novaes14,rassat14,axelsson15}.

Along with the cosmic signal of our interest, a CMB detector would register
all the microwave emission including emission from our own galaxy and extragalactic sources
that compromises the CMB observations. As such the raw maps are \emph{cleaned}
using a variety of procedures \citep{tegmark03,eriksen04,leach08,saha11}.
Ideally the aim of every cleaning procedure
is to completely remove any trace of foregrounds from the raw data and obtain
the pristine cosmic signal. However in practice
due to high foreground emission in some regions of the sky including the galactic plane,
the recovered CMB signal still remains contaminated in those parts of the sky.
To avoid spurious inferences from the bias thus induced, for example in obtaining
the angular power spectrum or cosmological parameters there off, a cleaned CMB
map is masked omitting regions of the sky where the recovered CMB is unreliable.
The quantities of interest are thus estimated from the unmasked region.

In this paper, we study a clean CMB map for foreground residuals that may still
be present in it, even after employing a foreground mask.
Since the foreground emission is strong in some regions of the CMB sky,
a suitable mask (a common mask or a frequency specific mask) is devised to
eliminate pixels beyond a chosen threshold level \citep{w1yrfg,tegmark03}.
No significant foreground residuals are expected to be present outside a mask
such as the $KQ75$ mask used in the NASA's WMAP nine year data analysis \citep{w9yrmaps}.
Further, we will probe any anomalous regions thus identified, if they were
earlier reported in the CMB literature and thus have already drawn
some interest.

Here we make use of a cross-correlation statistic to probe foreground residuals,
that quantifies the level of correlation between a clean CMB map and a foreground
template. This correlation coefficient is computed outside a galactic mask.
A cross-correlation analysis of CMB data with external data sets was already used
in the literature in a variety of ways, for example, in \citet{afshordi04,coles04,dineen04,bernardi05,
naselsky05,alvarez06,rassat07,land07,sarkar09,monteagudo10,corredoira10,naselsky10,
sawangwit10,ilic11,taburet11,hansen12,WW15}.

The paper is organized as follows.
In section~\ref{sec:stat}, we will introduce the cross-correlation statistic
used for probing foreground residuals that may still be present outside a
chosen mask. Then in section~\ref{sec:datasim}, the data sets used for
foreground templates, clean CMB map and foreground mask, and preparation
of simulations to complement the observed data, are discussed. In
section~\ref{sec:anls}, the statistic is applied to the data and regions that are
significantly correlated with the foreground templates are mapped out. Further
we carry out a local analysis of regions of anomalous correlation, that were
found in earlier studies and have received significant attention.
Finally a brief discussion and conclusions are presented in section~\ref{sec:concl}.

\section{Cross-correlation statistic}
\label{sec:stat}
We employ a cross-correlation coefficient (CCC) that would \emph{map} the correlations
between a clean CMB map and a foreground template in small patches over the entire sky.
It is defined to identify regions containing anomalous foreground
residuals outside a foreground mask.
Using the \texttt{HEALPix} sky discretization scheme, we define the CCC as \citep{alurithesis}
\begin{equation}
\mathcal{R}(P) = \frac{\sum_{p \in P}\left[T(p)-\bar{T}(P)\right]\left[F(p)-\bar{F}(P)\right]}
   {\sqrt{\sum_{p \in P}\left[T(p)-\bar{T}(P)\right]^2\sum_{p \in P} \left[F(p)-\bar{F}(P)\right]^2}}\,,
\label{eq:ccc-stat}
\end{equation}
where $P$ corresponds to a pixel in a lower resolution (lower $N_{side}$) \texttt{HEALPix}
grid that is used as mask on a high resolution (higher $N_{side}$) CMB map,
$T(p)$, and foreground template, $F(p)$, whose pixels are indexed by $p$.
So, the correlation coefficient between a CMB map and a foreground is computed
using all the pixels $p$ in the input maps,
falling in the region covered by the upgraded larger pixel from a lower $N_{side}$
\texttt{HEALPix} grid (upgraded to the same \texttt{HEALPix} resolution as input maps).
The CCC thus computed is associated with the pixel
$P$ of the low resolution \texttt{HEALPix} grid to obtain an all sky map of
the correlations between cleaned CMB data and foregrounds.
The local means, $\bar{T}(P)$ and $\bar{F}(P)$, of the CMB map
and foreground templates are also obtained from the pixels $p \in P$ of the
same larger pixel region from the low resolution \texttt{HEALPix} grid at which we
are mapping the correlations.

To illustrate, for input clean CMB and foreground maps at \texttt{HEALPix}
resolution of $N_{side}=512$, we may choose a lower resolution
\texttt{HEALPix} grid of $N_{side}=16$ to robustly map the correlations between the
input maps. So, the CCC is computed using a total of $(512/16)^2=1024$ pixels
of the input maps that fall into the same large pixel region corresponding
to the low resolution \texttt{HEALPix} $N_{side}=16$ grid.

Since we would map the entire sky, we downgrade the high resolution foreground
mask with $\{1,0\}$'s to the same $N_{side}$ as the CCC map. Further we exclude
the pixels which are not \emph{one}, meaning pixels that are at the edge of the
mask and didn't form a whole pixel at the lower resolution are excluded, before
applying it to CCC map.

However, since a \texttt{HEALPix} tessellation is made of disjoint pixels,
some of the anomalous regions may be cut-in, and would return a cross-correlation
coefficient that spuriously depicts it as insignificant. Therefore, we use a circular
disk of a chosen radius centered at each pixel of a chosen \texttt{HEALPix} grid
to continuously scan the input maps and compute the CCC. It is then associated
with that central pixel about which the circular disc mask is defined
to obtain a CCC map between CMB and foregrounds.
The CCC map is obtained at a lower resolution than the input
maps for convenience of computation but nevertheless maps the correlations sufficiently
smoothly. This circular disc is taken in union with a foreground mask so that
pixels in the region circumscribed by the circular disc which are flagged as
contaminated by the foreground mask does not contribute to the CCC statistic being
computed. Such a union will result in less number of pixels being available for computing
the statistic.

In order to have a robust estimate of the CCC statistic, we compute
the CCC only if atleast $80\%$ of the pixels are retained in the union of the foreground
mask and circular disc centered at any pixel of the chosen \texttt{HEALPix} grid
for obtaining the CCC map. The local means $\bar{T}(P)$ and $\bar{F}(P)$
are also computed using only the unmasked pixels in the effective circular disc.
If the $80\%$ criteria is not satisfied in a local effective circular mask
at any sky location compared to the full disc, it is deemed invalid. In practice, we set these invalid
pixels to \texttt{HEALPix} bad/missing pixel value of $-1.6375\times10^{30}$.
Further we can take the means $\bar{T}$ and $\bar{F}$ to correspond to those
of the respective full maps outside the foreground mask, so that we would be
considering deviations with respect to the full map means.
Thus our statistic of Eq.~[\ref{eq:ccc-stat}], would be same as that of \citet{hansen12}.
In this paper, we would be using this version of the statistic.

Again, to illustrate, for an input CMB map and a foreground template at $N_{side}=512$,
we can compute the CCC map at $N_{side}=128$, for example, to smoothly map the correlations
between them. Using a circular disc mask of radius $5^\circ$ (degrees), for example, at each sky
location in union with a foreground mask, we would be using atleast
$\approx 4792$ pixels of the input maps wherever available, as explained later.
If these many are not available, the corresponding pixel of CCC map is set to
\texttt{HEALPix} bad value.

\section{Data sets and simulations}
\label{sec:datasim}

\subsection{Observational data used}
In this work, we use the WMAP's nine year internal linear combination map (WILC9 hereafter) \citep{eriksen04,w9yrmaps} as the cleaned CMB map to probe foreground residuals.
As described later, we would be using a sufficiently extended
mask to omit regions where the recovered CMB signal is unreliable.
So, CMB sky obtained using any of the cleaning procedures are expected to be
in agreement with each other outside a foreground mask.
The WILC9 map is available at \texttt{HEALPix} $N_{side}=512$ with a resolution
of Gaussian beam of $FWHM=1^\circ$.

For galactic foreground templates, we use the thermal dust template at $94$~GHz
\citep{fdsmap}, $H\alpha$ map that is used as proxy for galactic free-free emission \citep{halphamap},
and the $408$~MHz Haslam map dominated by synchrotron emission from our galaxy \citep{haslammap}.
These independent foreground emission templates are used by WMAP as well as PLANCK teams as priors
in their foreground component estimation/separation
methods\footnote{All these three foreground templates are publicly available
from the NASA's Legacy Archive for Microwave Background Data Analysis (LAMBDA) web page : \url{http://lambda.gsfc.nasa.gov/}.} \citep{w9yrmaps,plk15fg}.
Hereafter we refer to these foreground templates as FDS, $H\alpha$ and Haslam maps
respectively.
Eventually, we will also make use of the foreground maps obtained
using the CMB observations from ESA's PLANCK satellite to study
local anomalous regions. We specifically use the galactic thermal dust, free-free and
synchrotron maps estimated using the \texttt{Commander}\footnote{\url{http://commander.bitbucket.org/}}
algorithm that are part of the PLANCK's 2015 data release\footnote{PLANCK's public release~2 data is
available at \url{http://irsa.ipac.caltech.edu/Missions/planck.html}.} \citep{plk15fg}.
As these foreground maps are obtained from a completely different mission
with different noise and systematics, they would complement the three foreground
templates mentioned above, unlike the MEM or MCMC derived foreground maps
from WMAP data itself\footnote{\url{http://lambda.gsfc.nasa.gov/}} \citep{w9yrmaps}.

Since the original \texttt{HEALPix} resolution and beam smoothing of the individual maps
are different, we downgrade them to a common \texttt{HEALPix} $N_{side}=256$ and
to have a resolution of $1^\circ$ FWHM Gaussian beam.
All the analysis presented in this work is carried out using data maps,
foreground mask and simulations at $N_{side}=256$. The CCC
map and it's significance map are obtained at $N_{side}=128$.

\begin{figure}
\centering
\includegraphics[width=0.47\textwidth]{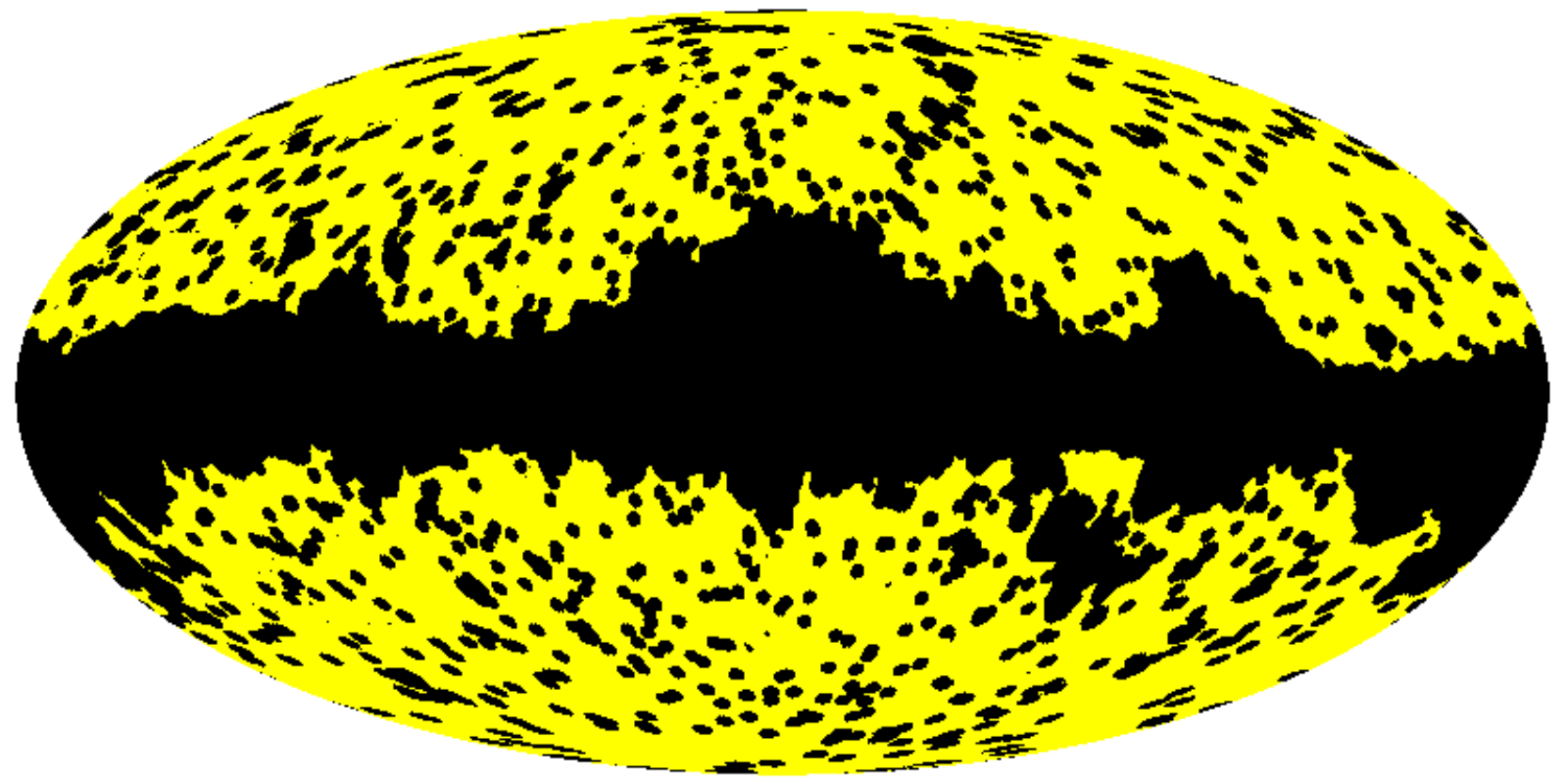}
\caption{Shown here is the foreground mask we use in the present analysis
         which is at $N_{side}=256$ and has an available sky fraction of $f_{sky} \approx 0.530$.
         Yellow and black regions denote the unmasked and masked parts of the sky, respectively.
         It is obtained by suitably processing and extending the $KQ75$ mask used by the
         WMAP science team in their nine year analysis. See text for details.}
\label{fig:kq75nside256}
\end{figure}

As a foreground mask, we use the WMAP's nine year $KQ75$ extended temperature mask
which has an unmasked sky fraction of $f_{sky} \approx 0.688$. However, the $KQ75$ mask
is available at $N_{side}=512$. In order to
obtain a suitable mask at $N_{side}=256$ where we perform all the analysis, the
$KQ75$ mask is first inverted ($\{1,0\}\rightarrow\{0,1\}$ respectively).
The inverted mask is then downgraded to $N_{side}=256$, and smoothed with a Gaussin beam
of $FWHM=1^\circ$ (degree) which is same as the beam resolution of the observed maps.
A cutoff of $0.01$ is imposed
on the inverted, downgraded, smoothed mask so as to extend it to handle leakage of
foregrounds due to smearing of extended and point sources due to beam smoothing and
downgrading of the observed maps which are available at higher $N_{side}$.
Some of the island pixels are also removed before inverting
it to get a reliable $\{1,0\}$ foreground mask at $N_{side}=256$. The mask thus obtained
and used throughout our analysis is shown in Fig.~[\ref{fig:kq75nside256}],
which has an unmasked sky fraction of $f_{sky} \approx 0.530$.

\subsection{Preparing the simulations}
To complement the observed CMB maps and compute significances of
the observed correlations, we generate an ensemble of WILC9-like
maps. We do so as follows.

First, we simulate a set of $1000$ isotropic, Gaussian CMB realizations
based on a best fit theoretical angular power spectrum ($C_l$). The best fit
$C_l$ are obtained using WMAP's nine year cosmological parameters \citep{w9yrcosmo}
as input to Code for Anisotropies in the Microwave
Background\footnote{The code is available for public download at \url{http://camb.info/}.}
(CAMB) \citep{camb1,camb2}.
Each random realization is convolved with beam transfer functions ($b_l$) corresponding
to each of the five WMAP channels\footnote{\url{http://lambda.gsfc.nasa.gov/}} to
create five beam smoothed maps.
These convolved maps are then added with anisotropic frequency specific detector noise
simulated using the noise rms ($\sigma_0$) corresponding to each observed frequency band map,
and the effective number of observations in each pixel/sky direction ($N_{obs}(p)$).
The effective noise rms in a pixel `$p$' of the measured anisotropies
at a frequency band `$i$' is given by $\sigma^i(p)=\sigma^i_0/\sqrt{N^i_{obs}(p)}$.
The $\sigma_0$ values and $N_{obs}$ maps are also provided by the WMAP team with the
nine year data release\footnote{\url{http://lambda.gsfc.nasa.gov/}}. Thus we simulate
random Gaussian noise maps using $\sigma^i(p)$ corresponding to the WMAP's
$23(K)$, $33(Ka)$, $41(Q)$, $61(V)$ and $94(W)$~GHz channels.

In this way, five sets of 1000 mock WMAP nine year observed maps were generated with
appropriate channel specific beam and noise properties. To obtain
an ILC-like map, we deconvolve each of the mock channel maps with the
respective beam window functions, and smooth them with a Gaussian
beam of $FWHM=1^\circ$. Now, we co-add the smoothed WMAP-like channel maps using
smoothed noise variance maps to obtain WILC9-like maps.
So, we treated the covariance matrix to be essentially diagonal, and used the
smoothed noise variance maps themselves as weights for taking the linear
combination\footnote{Alternatively, one can use the same weights used for linearly combining
the WMAP nine year smoothed raw maps to get the ILC CMB map. This should be done iteratively over
all the 12 sky partitions in the same way as the observed maps were combined to obtain a
composite clean CMB map. Since the simulations used are independent random realizations
of the observed sky, the statistical independence of the resultant ILC-like maps would still be intact,
despite using fixed weights for linearly combining them. This way we can bypass the ambiguity
over the uncertainty regarding the use of inverse variance weighing to combine the smoothed
band maps, rather than using the inverse of the full noise covariance matrix.
Further we will be truly mimicking the iterative cleaning employed in ILC procedure.
The iterative cleaning is used to deal with different levels of foreground emission in various
parts of the CMB sky by choosing appropriate weights suitable for that region of the sky.}.
These maps are then downgraded to $N_{side}=256$.
Thus we generated a set of $1000$ ILC-like CMB maps at $N_{side}=256$.
In the next section, we proceed with the analysis of observed maps and estimate
significances using these simulations.

\section{Analysis and results}
\label{sec:anls}

\subsection{Foreground residuals in a clean CMB map}
In this section, we probe the presence of residual foregrounds, if any, in the
WMAP's nine year ILC cleaned CMB map (WILC9) by cross-correlating it with the foreground
templates of thermal dust, free-free and synchrotron emission from our own galaxy.

We apply our CCC statistic defined in Eq.~[\ref{eq:ccc-stat}]
to obtain a CCC map with each foreground at $N_{side}=128$. The
input WILC9 map and a foreground template at $N_{side}=256$ are scanned continuously using circular
discs of radii $1^\circ$, $2^\circ$, $3^\circ$, $4^\circ$, $5^\circ$ and $6^\circ$ (degrees).
This is done in union with the foreground mask at $N_{side}=256$ derived from WMAP's nine year
$KQ75$ mask, with the additional constraint of having atleast
$80\%$ of usable pixels in the effective circular mask, compared to the full disc mask of a
chosen radius. This is to ensure a robust computation of the CCC at a chosen location on the sky.
By using a circular disc of radius $1^\circ$, and demanding availability
of atleast a fraction of $0.8$ number of pixels in the net disc mask after combining
with the foreground mask, we will be using atleast $\approx 48$ pixels for computing
the CCC at any location of the sky (treating each pixel of an $N_{side}=256$ map as a square,
it would have a size of $a=\sqrt{4\pi/(12\times256^2)}\times 180^\circ\,60'/\pi=13.74'$ (arcmin)
resulting in demanding availability of atleast $0.8 \times \pi r^2_{disc}/a^2 \approx 47.9$ pixels).
Likewise we will be using atleast $\approx 192$, $431$, $767$, $1198$ and $1725$
pixels to compute CCC when using circular discs of $2^\circ$, $3^\circ$, $4^\circ$, $5^\circ$ and
$6^\circ$ (degrees) radii, whereever the desired constraint is satisfied locally on the sky.

\begin{figure*}
\centering
\includegraphics[width=0.56\textwidth]{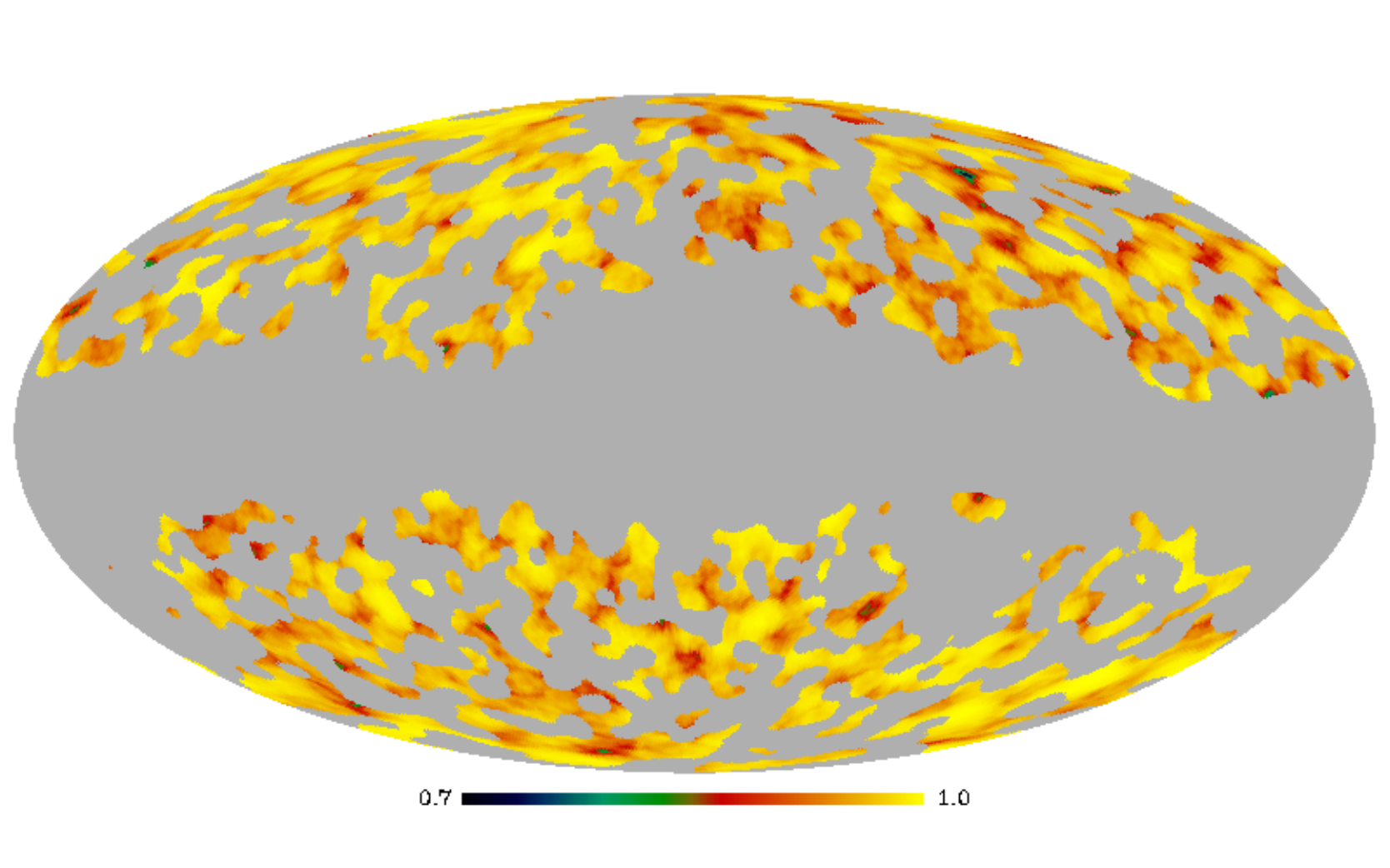}
~
\includegraphics[width=0.38\textwidth]{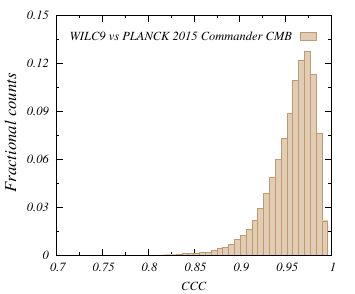}
\caption{\emph{Left:} The correlation map of clean CMB maps obtained from two
         different missions viz., WMAP's nine year ILC map (WILC9) and the PLANCK's 2015
         \texttt{Commander} CMB map, using a filter disc radius of $3^\circ$ (degrees)
         at $N_{side}=128$ is shown here. As expected both of them are highly
         and positively correlated across the sky.
         \emph{Right:} Histogram plot of all the correlation coefficients from
         the CCC map shown to the \emph{left}. We readily see that the two maps
         are well correlated as expected. A total of $62721$ pixels are valid,
         out of the $12\times128^2=196608$ pixels of an $N_{side}=128$ map
         ($\approx 32\%$ of the sky), in the CCC map shown to the left.
         The histogram plot is made by sorting these valid pixels into $50$ bins.}
\label{fig:wilc9plk15}
\end{figure*}

To begin with the analysis, we test the level of correlation between two clean CMB
maps viz., WMAP's nine year ILC map and the \texttt{Commander}
CMB map from PLANCK's 2015 data release \citep{eriksen08,plk15cmb}.
Their cross-correlation map is shown in the \emph{left} plot of Fig.~[\ref{fig:wilc9plk15}].
We chose a filter disc radius of $3^\circ$ to obtain this CCC map at $N_{side}=128$.
Also shown in Fig.~[\ref{fig:wilc9plk15}] (\emph{right}) is a histogram of the pixels
of this CCC map. It is readily evident that the two clean CMB maps from different missions
are highly correlated, as expected, with most of the coefficients across the sky closer
to `one'. Since we take the union of a circular disc with the foreground mask shown
in Fig.~[\ref{fig:kq75nside256}], with a further constrain of having atleast $80\%$ of
the pixels in the net circular disc, the net sky over which CCC is computed is smaller
that the foreground mask itself ($\approx 32\%$ of the whole sky). The grey pixels
didn't satisfy these constraints and are left out of the analysis. Thus, if any
foreground residuals are found eventually, we are warranted of (still) using
the WILC9 map to probe foreground residuals in cleaned CMB data.

\begin{figure*}
\centering
\includegraphics[width=0.32\textwidth]{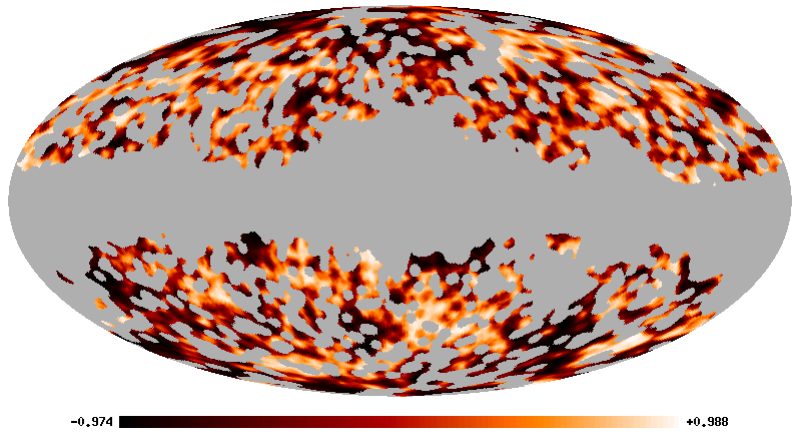}
\includegraphics[width=0.32\textwidth]{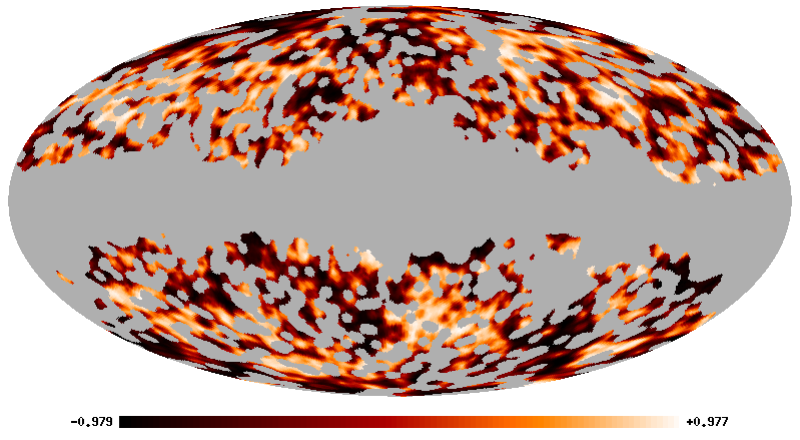}
\includegraphics[width=0.32\textwidth]{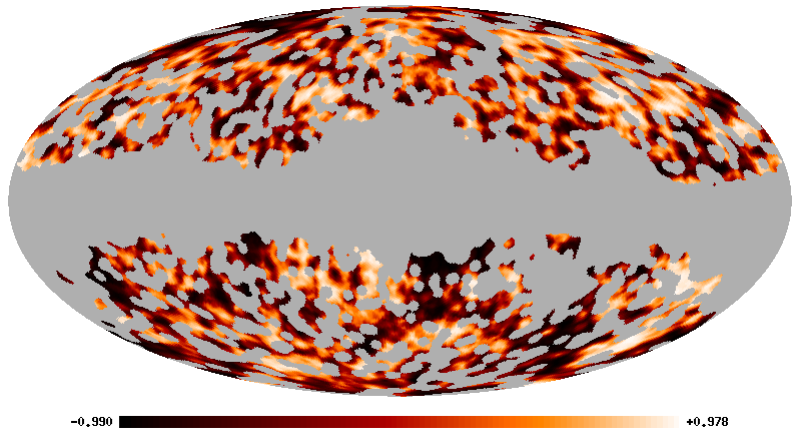}
\includegraphics[width=0.32\textwidth]{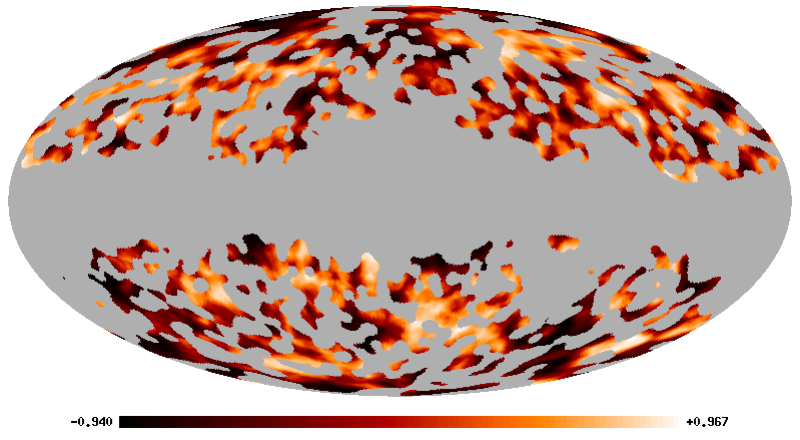}
\includegraphics[width=0.32\textwidth]{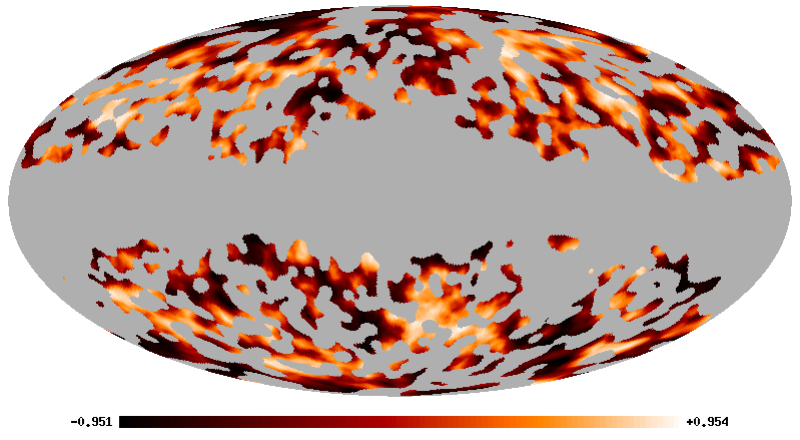}
\includegraphics[width=0.32\textwidth]{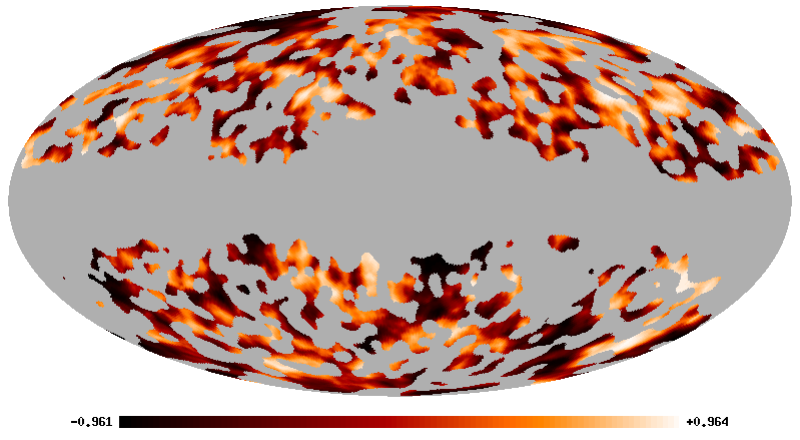}
\includegraphics[width=0.32\textwidth]{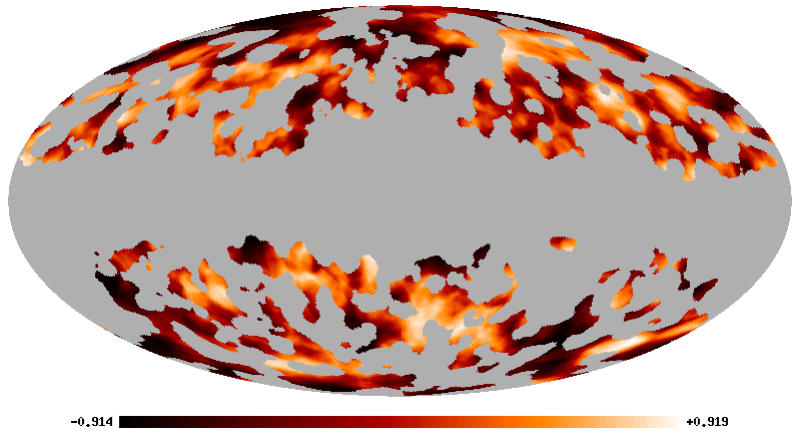}
\includegraphics[width=0.32\textwidth]{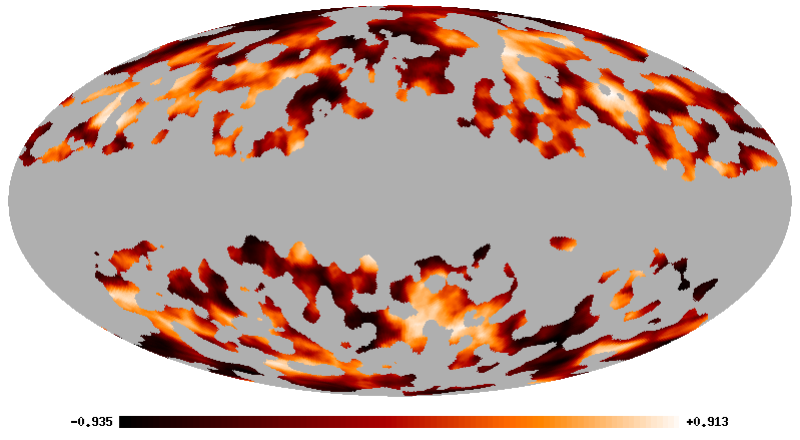}
\includegraphics[width=0.32\textwidth]{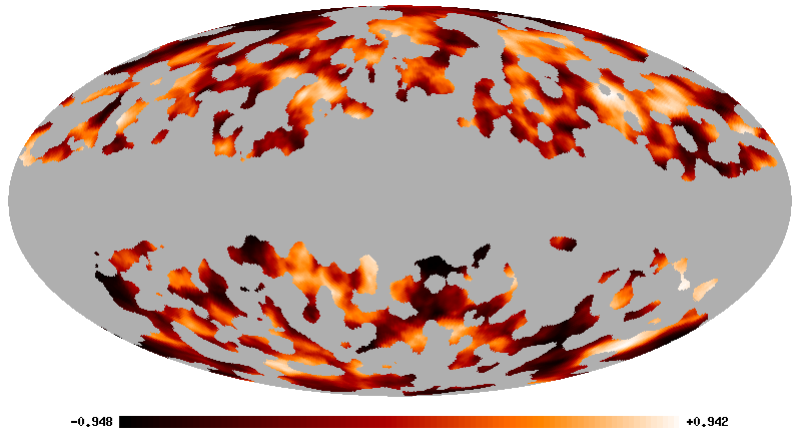}
\includegraphics[width=0.32\textwidth]{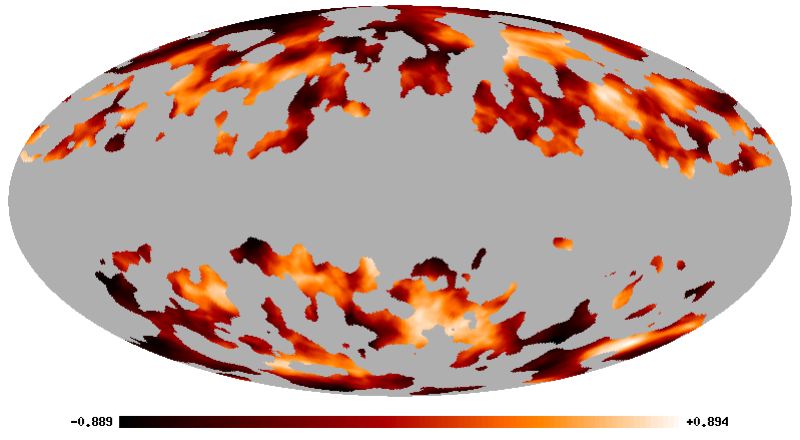}
\includegraphics[width=0.32\textwidth]{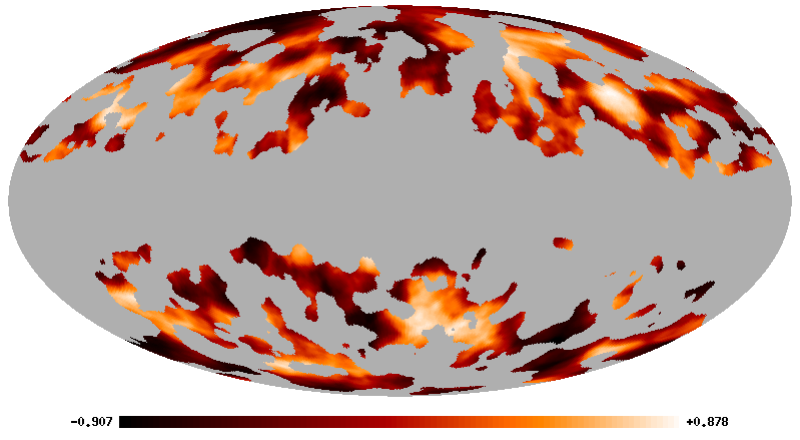}
\includegraphics[width=0.32\textwidth]{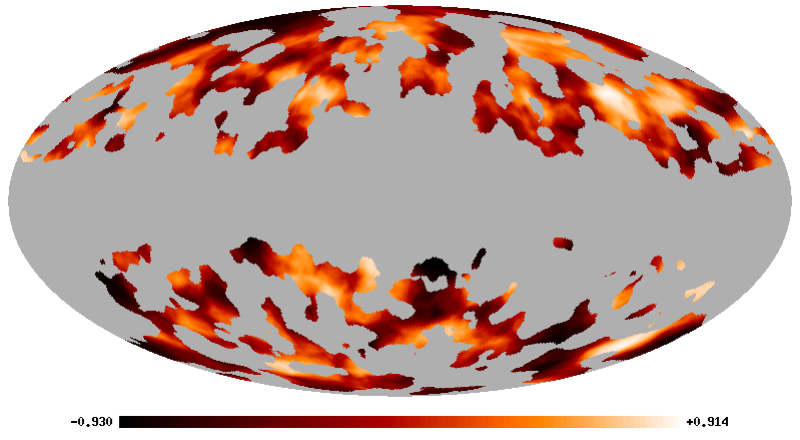}
\caption{Correlation maps of WMAP's nine year ILC map with the foreground templates
         FDS thermal dust map (\emph{first column}), H$\alpha$ map (\emph{second column}),
         and the Haslam's $408$~MHz map (\emph{third column}) are presented here. The CCC
         maps obtained using circular disc masks of radii $2^\circ$, $3^\circ$, $4^\circ$,
         and $5^\circ$ (degrees) are shown from \emph{first} to \emph{fourth} row, respectively.}
\label{fig:cccdata}
\end{figure*}

Moving towards the main analysis, we now correlate the WILC9 clean CMB map with
foreground templates viz., FDS map for thermal dust, $H\alpha$ for free-free,
and Haslam map for synchrotron emissions from our own galaxy.
The input CMB and foreground maps are at \texttt{HEALPix} $N_{side}=256$, and the
output correlation maps are obtained at $N_{side}=128$.
The resulting CCC maps are shown in Fig.~[\ref{fig:cccdata}]. Circular disc
masks of $1^\circ$ to $6^\circ$ (degrees) radii are used at each sky position,
in steps of one degree, to filter the residual foregrounds by angular size.
However the results are shown corresponding to the disc masks of size $2^\circ$ to
$5^\circ$ (degrees) radii only, in Fig.~[\ref{fig:cccdata}].
The local circular disc masks are used in conjugation with the suitably modified
$KQ75$ mask from WMAP nine year data shown in Fig.~[\ref{fig:kq75nside256}].
By demanding the availability of atleast
$80\%$ of pixels in the net circular mask compared to the full disc at any location
on the sky, the correlations are mapped over
a fraction of $\approx 0.45$, $0.38$, $0.32$, $0.31$, $0.30$ and $0.29$ of the sky,
for our chosen disc masks of radii $1^\circ$, $2^\circ$, $3^\circ$,
$4^\circ$, $5^\circ$ and $6^\circ$ (degrees) respectively.

\begin{figure*}
\centering
\includegraphics[width=0.32\textwidth]{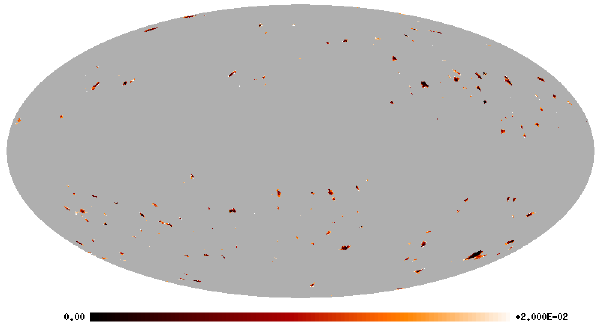}
\includegraphics[width=0.32\textwidth]{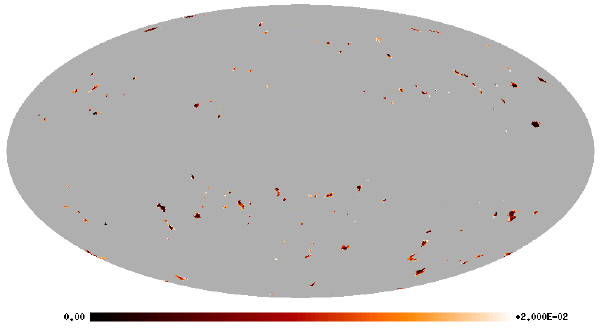}
\includegraphics[width=0.32\textwidth]{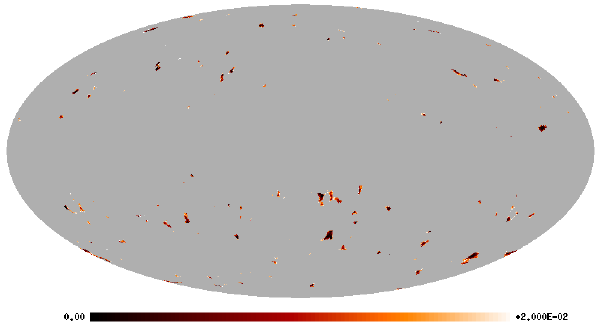}
\includegraphics[width=0.32\textwidth]{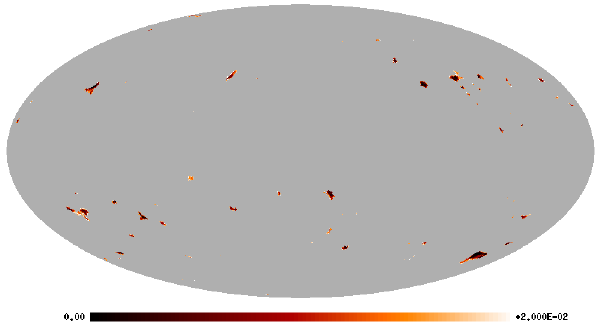}
\includegraphics[width=0.32\textwidth]{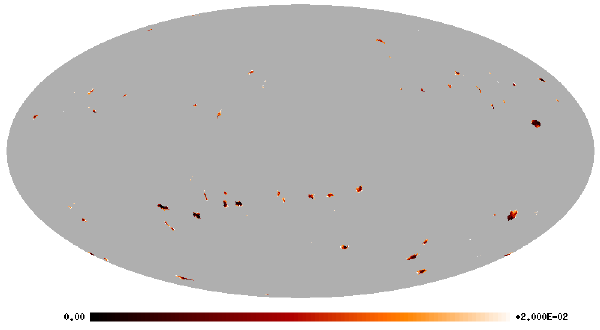}
\includegraphics[width=0.32\textwidth]{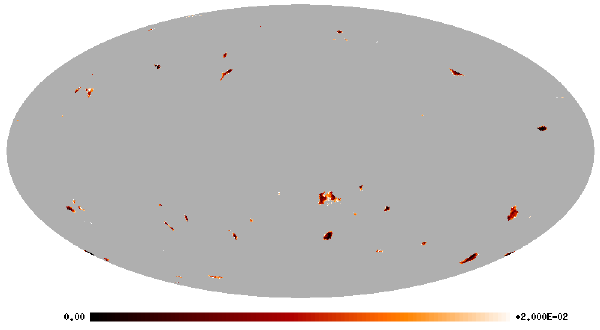}
\includegraphics[width=0.32\textwidth]{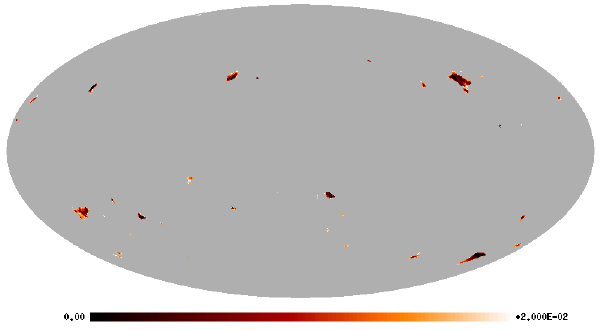}
\includegraphics[width=0.32\textwidth]{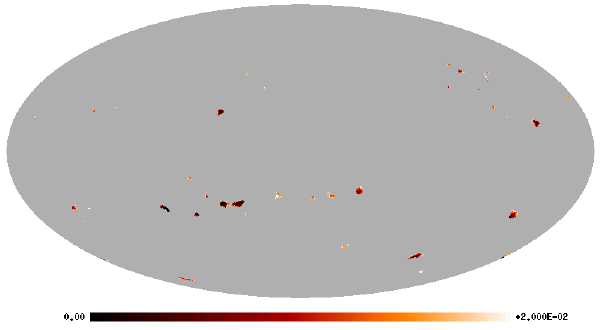}
\includegraphics[width=0.32\textwidth]{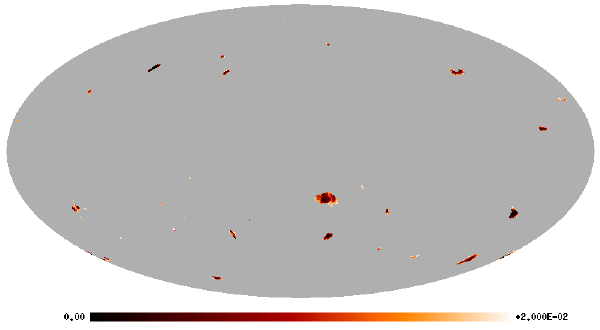}
\includegraphics[width=0.32\textwidth]{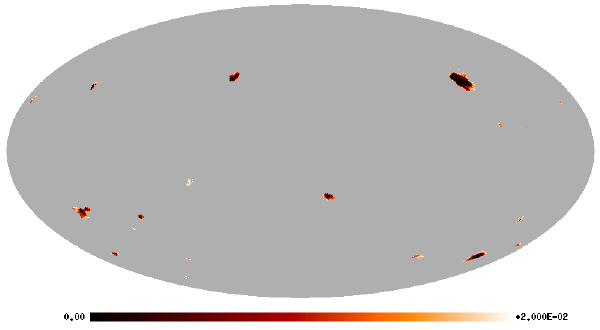}
\includegraphics[width=0.32\textwidth]{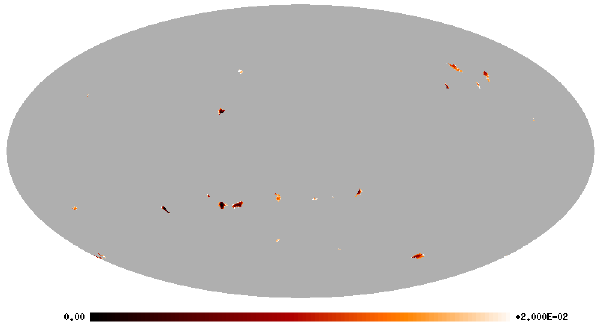}
\includegraphics[width=0.32\textwidth]{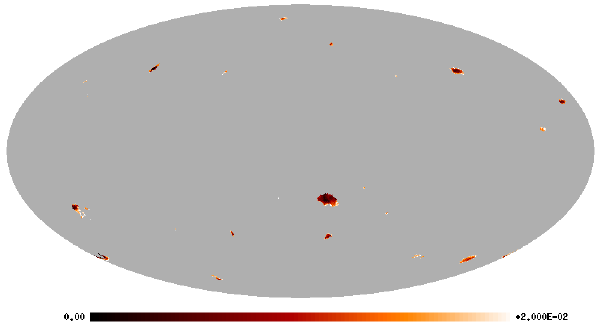}
\caption{$p-$value maps of correlations of WILC9 CMB map with foreground
         templates viz., FDS thermal dust, H$\alpha$, and the Haslam maps
         are shown here corresponding to those presented in Fig.~[\ref{fig:cccdata}].
         Only those regions which are anomalous with a \emph{probability to exceed}
         by $p=0.02$ i.e., a $2\%$ or less by a random chance are highlighted.
         One can readily see that the $p-$maps are almost empty.
         There are only few regions with significant residual contamination, as expected.}
\label{fig:pmaps}
\end{figure*}

One should be warned of directly drawing inferences from the CCC maps of Fig.~[\ref{fig:cccdata}],
as there are regions of apparent high correlation and anti-correlation between the clean
CMB map and foregrounds, as indicated by the color scaling used. 
In order to quantify their significance, we obtain $p-$maps corresponding to each of
the correlation maps between clean CMB data and foregrounds.
$p-$maps are maps of $p-$values for each valid pixel at any location of the sky at which
CCC is computed between WILC9 CMB map and foreground templates.
So, they too span the same sky region and have the same $N_{side}$ as the observed
correlation maps shown in Fig.~[\ref{fig:cccdata}].
The $p-$maps are obtained by correlating the foreground templates with the isotropic,
Gaussian simulations described in section~\ref{sec:datasim}. All the simulated $1000$
WMAP nine year ILC-like maps are correlated with the thermal dust, free-free and
synchrotron foreground templates, and the \emph{probability to exceed} the observed
correlation in the data at each valid pixel of the CCC maps is computed. This $p-$value
is then associated with a pixel of same index to obtain a $p-$map corresponding to a
data CCC map.
However, to identify the anomalous regions with significant foreground contamination,
we chose a cutoff $p-$value exceeding which we retain that pixel in the $p-$map.
The $p-$maps thus obtained are shown in Fig.~[\ref{fig:pmaps}] for a cutoff $p-$value
of $0.02$ (i.e., a random chance occurrence probability of $2\%$ or less).

As expected, the $p-$maps are largely empty. However they reveal that there are still
few extended regions of significant contamination. With smaller disc mask radius there are
many stray or small island pixel regions which were found to be anomalous along with
some extended patches. But, using larger discs for probing the residuals,
only relatively extended or blob like regions remain. In Fig.~[\ref{fig:pmaps}],
we have again shown only the $p-$maps 
for circular discs of radii $2^\circ$ to $5^\circ$ (degrees), corresponding to those
of Fig.~[\ref{fig:cccdata}]. With $1^\circ$ (degree) radius circular disc, there are many stray
and few pixel islands, which may be arising due to it's smaller disc size
that allows the use of only $\pi (60')^2/(13.74')^2 \approx 60$ pixels at best to compute
the CCC locally. Among these already few pixels there may be an unresolved point source or some may be
sitting on the edge of the foreground mask used.
Using disc of angular size $6^\circ$, the anomalous regions are similar to those
from using the $5^\circ$ circular mask, if not for some stray pixels/islands. Hence
these results are not shown for brevity.

We also checked the $p-$maps with a cutoff of $0.01$ (i.e., $1\%$ or less probability
to exceed the observed data correlation). Except for the disappearance of stray/island pixels,
the $p-$value maps essentially look the same. The same extended regions still
remain anomalously correlated with the foregrounds.

We notice that, with disc masks of larger radii, extended regions of anomalous
correlation are found in CCC maps with thermal dust and synchrotron templates and,
not so, relatively, with the free-free template. This may be due to the fact that
free-free does not dominate at any of the frequency channels, in the same way as
the galactic synchrotron and thermal dust emissions, in which WMAP made the observations.
Next, we will locally analyze interesting extended regions of anomalous correlation.

\subsection{Local analysis of anomalously correlated regions : CMB cold spot}
In this section, we locally probe the extended anomalous regions with
significant foreground correlation found in the previous section.
Among the identified regions with spurious foreground contamination,
we study the \emph{cold spot} region. Cold spot is an anomalously low
temperature region found in CMB sky in the direction $(l,b)=(209^\circ,-57^\circ)$,
in galactic co-ordinates, with an angular size of $\approx 10^\circ$ \citep{vielva04}.
Since it's detection, the CMB cold spot received significant attention
in the literature \citep{tomita05,cruz06,inoue07,rudnick07,cruz08,bernui09,naselsky10,
smith10,zhnag10,afshordi11,w7yranom,plk13anom,plk15anom}.

\begin{figure*}
\centering
\includegraphics[width=0.24\textwidth]{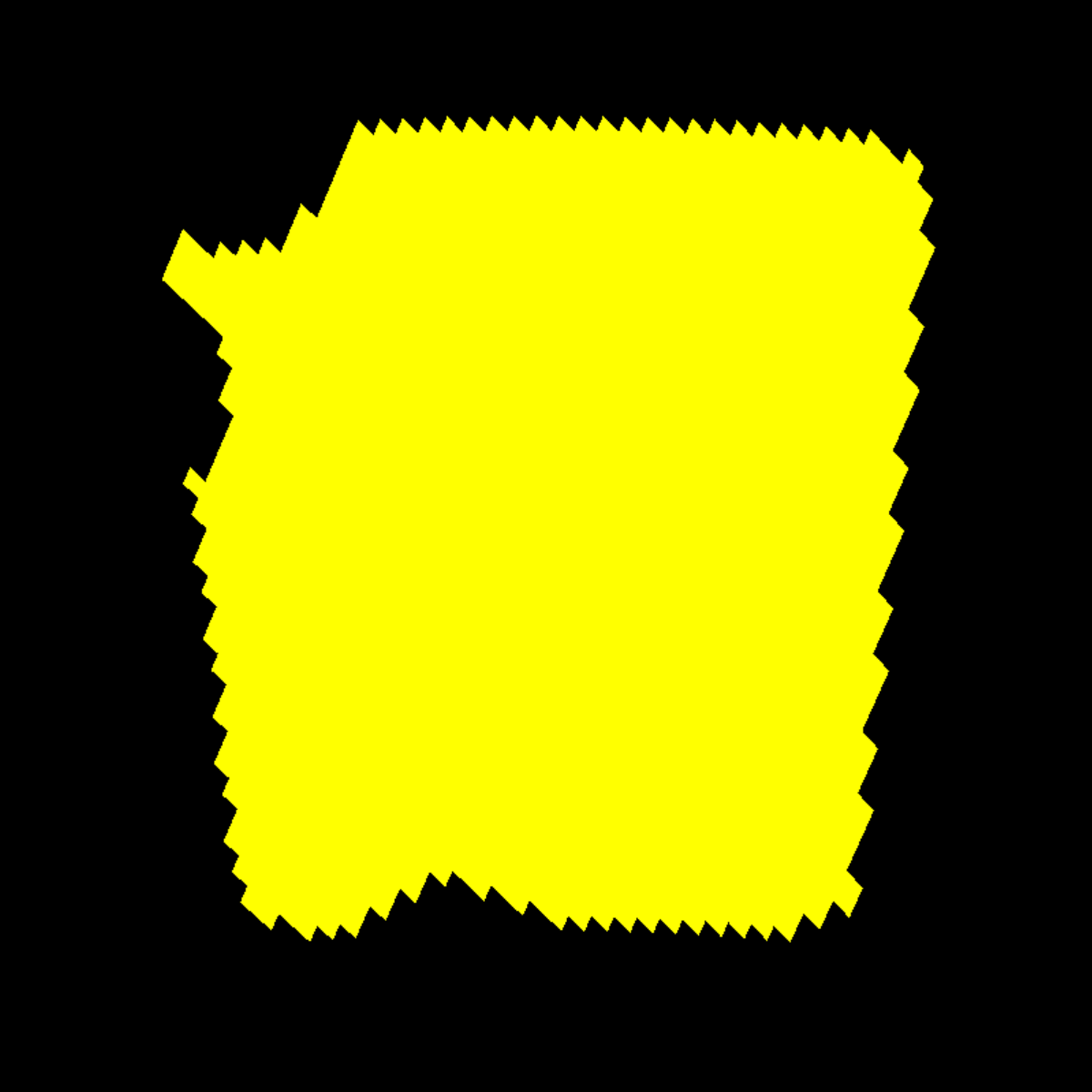}
~
\includegraphics[width=0.22\textwidth]{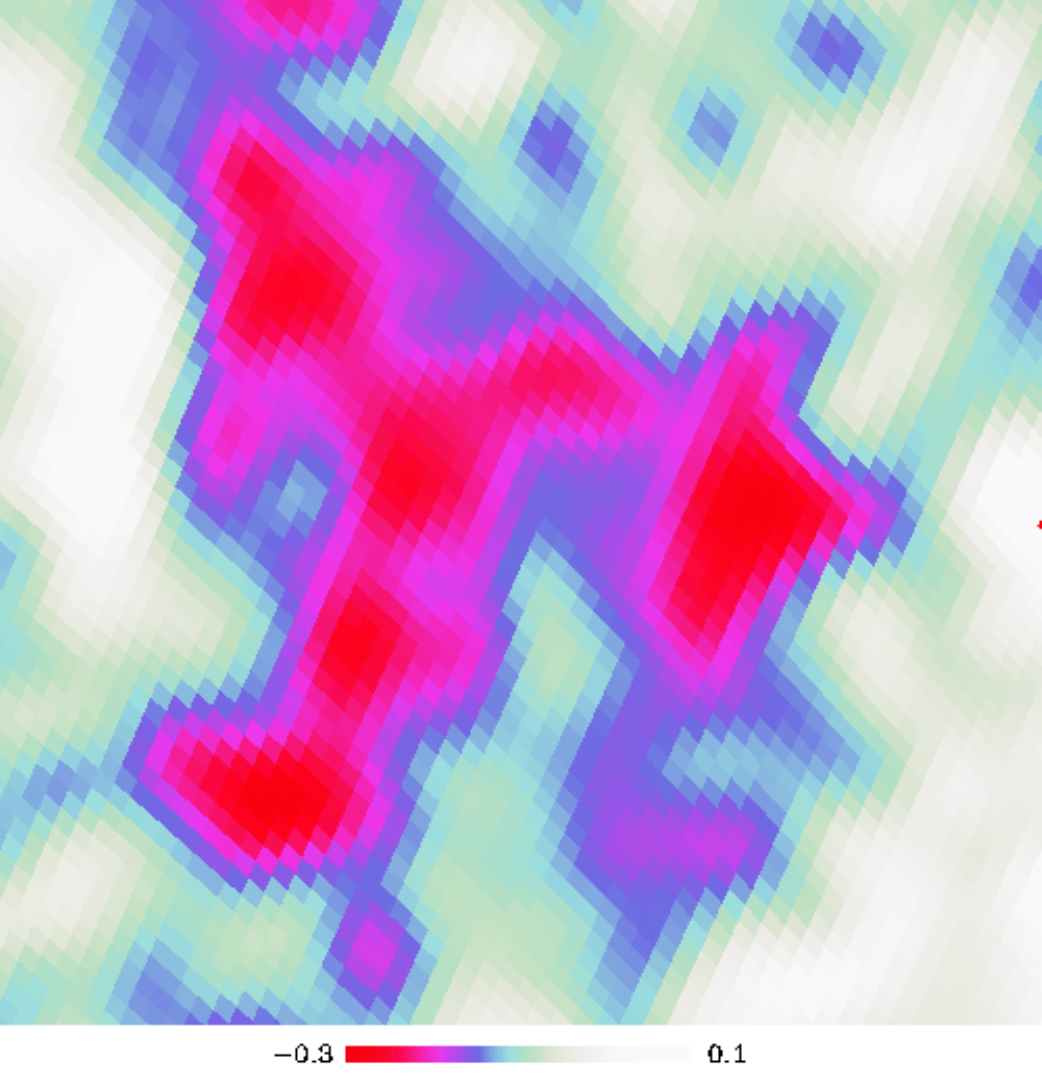}
~
\includegraphics[width=0.22\textwidth]{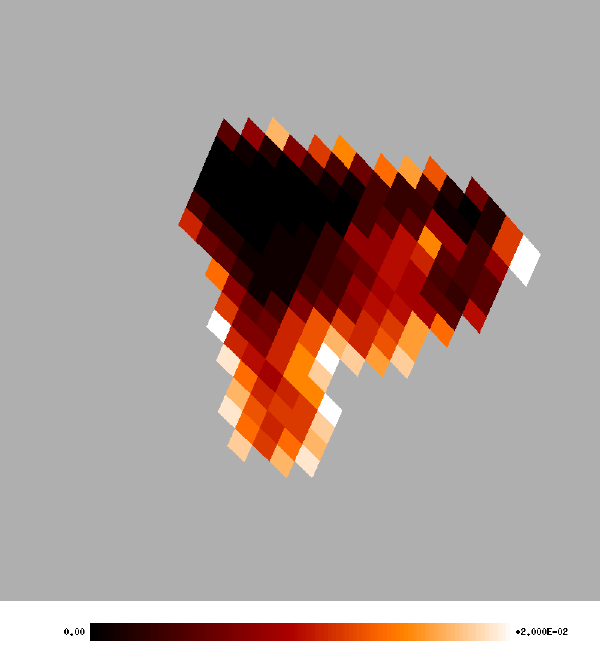}
~
\includegraphics[width=0.22\textwidth]{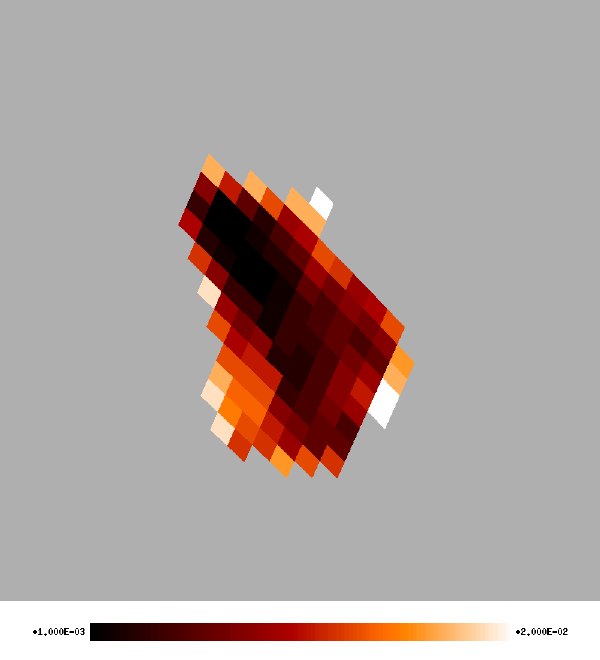}
\caption{\emph{First:} A square mask of size $10^\circ\times14^\circ$, that is used to
         further probe the local anomalous region of CMB cold spot, is shown here.
         Yellow and black denote unmasked ($1$) and masked ($0$) regions of the square mask.
         \emph{Second:} The CMB cold spot region from the WMAP's nine year ILC map at $N_{side}=256$
         is shown here. The extended pink feature (approximately $10^\circ$ in angular size
         in the direction $(l,b)=(209^\circ,-57^\circ)$, in galactic coordinates) 
         is the cold spot region.
         \emph{Third and Fourth:} The $p-$maps of CMB correlation with the thermal dust and
         synchrotron templates, respectively, obtained using a local filter disc radius of $3^\circ$
         are presented here. They are same $p-$maps shown in Fig.~[\ref{fig:pmaps}], but are
         shown centered in the cold spot direction shown to the left.}
\label{fig:cs-mask-obs-pmap}
\end{figure*}

From the CCC analysis of the previous section, we find that the
cold spot region of WILC9 CMB map is well correlated with the
FDS (thermal dust) and $408$~MHz Haslam (synchrotron) maps.
So we probe the cold spot region locally using a $10^\circ\times14^\circ$
square mask, and estimate the CCC and it's significance from that
region as a whole. This square patch mask in union with the
foreground mask of Fig.~[\ref{fig:kq75nside256}] is shown in
Fig.~[\ref{fig:cs-mask-obs-pmap}] (\emph{left}).
Also shown in Fig.~[\ref{fig:cs-mask-obs-pmap}], are the CMB cold spot
region itself from WILC9 map (\emph{second}), and the $p-$map's of CMB correlation with FDS dust
and 408~MHz Haslam map (\emph{third} and \emph{fourth})
shown in Fig.~[\ref{fig:pmaps}], corresponding to using $3^\circ$ radius circular
disc masks but centered towards the cold spot direction.

Using this square patch mask, we compute the correlation coefficient
defined in Eq.~[\ref{eq:ccc-stat}], between CMB and three foreground
templates. For completeness, we include the
free-free template also in this local analysis. Complementing the
FDS thermal dust map, 408~MHz Haslam's synchrotron map, and $H\alpha$
map for free-free emission, we also make use of the all-sky foreground maps
from PLANCK 2015 data release, estimated using the \texttt{Commander}
algorithm\footnote{\url{http://irsa.ipac.caltech.edu/Missions/planck.html}}
\citep{plk15fg}.
Since they are derived from a completely different mission with
different systematics and noise, they would augment the independent
templates used so far, in a way not possible with the
MEM and MCMC estimated foreground maps from WMAP data itself \citep{w9yrmaps}.

\begin{figure*}
\centering
\includegraphics[width=0.95\textwidth]{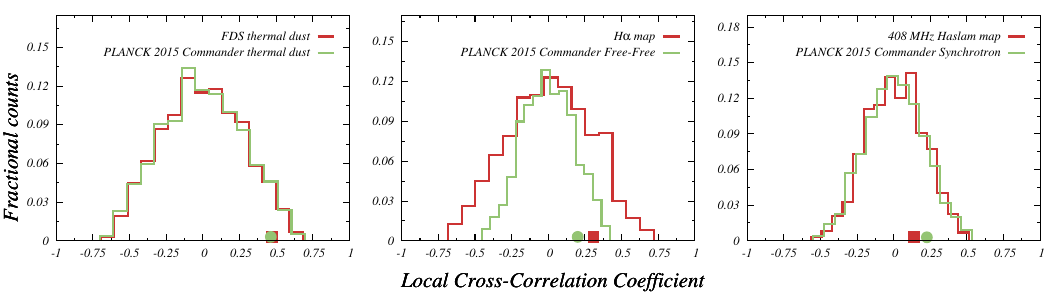}
\caption{Histogram plots corresponding to local correlation analysis of WILC9 map in the direction
         of cold spot region using the square patch mask shown in Fig.~[\ref{fig:cs-mask-obs-pmap}]
         with six foreground templates, is presented here. Independent foreground templates for
         galactic thermal dust (FDS map), synchrotron (Haslam map), and free-free
         ($H\alpha$ map) emissions, along with foreground maps provided with the
         PLANCK 2015 data release for the same three emission types are used here.
         The filled \emph{red square} and {green circle} denote observed CCC from the
         cold spot region when correlated with independent and PLANCK 2015 data derived
         foreground maps, respectively, used for each emission type.}
\label{fig:localccc}
\end{figure*}

The CCC from the cold spot region obtained using the square mask
is compared with $1000$ coefficients, computed by correlating $1000$ WILC9-like
CMB realizations with the three independent foreground templates as well
as with the PLANCK 2015 data derived foreground maps.
So, instead of a local circular disc mask as used in the previous section,
a fixed square mask is used to do
this local correlation analysis of cold spot region as a whole.
In Fig.~[\ref{fig:localccc}], we show the distribution of CCC from the
correlation of simulated CMB skies with the six foreground templates
in the cold spot region only, using the square mask.
The observed CCC values are also denoted, by a \emph{filled square} and
a \emph{filled circle} point types corresponding to two templates used
for each foreground emission type.
As expected we see that the CCC with $H\alpha$ map (\emph{middle} plot) is not
anomalous with a $p-$value$=0.172$.
The cold spot CCC of WILC9 CMB map with FDS thermal dust is found to
be anomalously correlated with a $p-$value=$0.046$, that is outside $95\%$
confidence level. However, the cold spot CCC of our clean CMB map with
Haslam map turns out to be insignificant with a $p-$value$=0.254$.
It is also readily evident that
the significances using PLANCK 2015 \text{Commander} estimated thermal dust,
free-free and synchrotron maps are similar to the CCC significances
using independent foreground templates viz., FDS, $H\alpha$ and Haslam maps.
The $p-$values of cold spot CCC with the PLANCK 2015 data derived
foregrounds are found to be $p=0.048,0.128,0.153$, respectively.

The significance found for the cold spot CCC of the clean CMB map with
synchrotron template is puzzling, despite what is suggested by $p-$map
from that region shown in Fig.~[\ref{fig:cs-mask-obs-pmap}] (\emph{fourth}).
From Fig.~[\ref{fig:cs-mask-obs-pmap}], the $p-$values contour for correlation
with synchrotron (\emph{fourth} figure) can be seen to be smaller than that
with thermal dust (\emph{third} figure). So, when we are using a
square patch mask of the same size to probe the cold spot region as a whole,
the significance of correlation with synchrotron is less because it contains less
anomalously correlated/contaminated pixels in it compared to the thermal dust.

\section{Conclusions}
\label{sec:concl}

In summary, in this paper we tried to investigate the presence of foreground
residuals that may still be present in a \emph{cleaned} CMB map after foreground
reduction. To that extent we carried out a correlation analysis of WMAP's nine
year ILC (WILC9) CMB map with three independent foreground templates corresponding to
the three dominant galactic emissions viz., thermal dust (FDS map), free-free
($H\alpha$ map) and synchrotron (408~MHz Haslam map).
The correlations are mapped out over the entire sky using a local circular disc mask
of different radii at each sky location to filter anomalous residuals by angular
size. The local circular disc masks are taken in union with the suitably modified
$KQ75$ temperature mask used in WMAP nine year data analysis.
This way we will be probing foreground residuals only in the supposedly clean regions
of the foreground reduced CMB map.
Cross-correlation coefficients (CCCs) are computed at pixels/sky directions
which have atleast $80\%$ of pixels in the net circular mask after
combining with foreground mask, compared to the full disc mask of a chosen radius.

The CCC maps of CMB with foreground templates are computed at $N_{side}=128$
with input CMB and foreground maps at $N_{side}=256$. Local circular
disc masks of different radii, viz., $1^\circ$, $2^\circ$, $3^\circ$, $4^\circ$, $5^\circ$ and $6^\circ$
(degree), are used to identify foreground residuals by angular size. A set of
$1000$ WMAP nine year ILC-like CMB skies are simulated to complement the data
used and find significances of the apparently well correlated regions with foregrounds.

Although, the CCC maps are largely empty, few regions with spurious correlations
were found. Some of the extended regions are persistent even with the largest
($6^\circ$ radius) circular disc mask used for filtering. A curious correlation
is found in the CMB cold spot region of WILC9 map with thermal dust and synchrotron templates.
With this finding, the cold spot region is further studied by performing a local correlation
analysis using a fixed square mask of size $10^\circ\times14^\circ$ encompassing the
cold spot region. The CCC of CMB with FDS thermal dust map from the cold spot
region as a whole using the square mask is found to be anomalous by more than $95\%$
confidence level. However the cold spot CCC with synchrotron is found to be insignificant
in this local analysis using square mask. It could be due to less number of contaminated
pixels in the chosen square region due to synchrotron compared to thermal dust.

This analysis may be used to include the few contaminated regions found here
with the usual foreground masks to make them more robust.

\section*{Acknowledgements}
We acknowledge the use of the Legacy Archive for Microwave Background Data Analysis (LAMBDA),
part of the High Energy Astrophysics Science Archive Center (HEASARC). HEASARC/LAMBDA is a
service of the Astrophysics Science Division at the NASA Goddard Space Flight Center.
Part of the results presented here are based on observations obtained with Planck
(\url{http://www.esa.int/Planck}), an ESA science mission with instruments and contributions
directly funded by ESA Member States, NASA, and Canada. We also acknowledge extensive use
of \textsc{HEALPix} software package in this work.

\label{lastpage}


\begin{thebibliography}{99}
\bibitem[\protect\citeauthoryear{Abramo, Sodre \& Wuensche}{2006}]{abramo06} Abramo~L.~R., Sodre~L. Jr., and Wuensche~C.~A., 2006, PRD, 74, 083515
\bibitem[\protect\citeauthoryear{Afshordi, Loh \& Strauss}{2004}]{afshordi04} Afshordi~N., Loh~Y.-S., and Strauss~M.~A., 2004, PRD, 69, 083524
\bibitem[\protect\citeauthoryear{Afshordi, Slosar \& Wang}{2011}]{afshordi11} Afshordi~N., Slosar~A., and Wang~Y., 2011, JCAP, 01, 019
\bibitem[\protect\citeauthoryear{Aluri}{2011}]{alurithesis}
Aluri~P.~K., \emph{Large scale anomalies in the Cosmic Microwave Background Radiation},
Ph.D. thesis submitted to Department of Physics, IIT Kanpur, India in December 2011
\bibitem[\protect\citeauthoryear{Aluri \& Jain}{2012}]{aluri12} Aluri~P.~K., and Jain~P., 2012, MNRAS, 419, 3378
\bibitem[\protect\citeauthoryear{Aluri et al.}{2011}]{aluri11} Aluri~P.~K., Samal~P.~K., Jain~P., and Ralston~J.~P., 2011, MNRAS, 414, 1032
\bibitem[\protect\citeauthoryear{Alvarez et al.}{2006}]{alvarez06} Alvarez~M.~A., Komatsu~E., Dore~O., and Shapiro~P.~R., 2006, ApJ, 647, 840
\bibitem[\protect\citeauthoryear{Axelsson et al.}{2015}]{axelsson15} Axelsson~M., Ihle~H.~T., Scodeller~S. and Hansen~F.~K., 2015, A\&A, 578, A44

\bibitem[\protect\citeauthoryear{Babich \& Pierpaoli}{2008}]{babich08} Babich~D., and Pierpaoli~E., 2008, PRD, 77, 123011
\bibitem[\protect\citeauthoryear{Bennett et al.}{2003a}]{wmap1yr} Bennett~C.~L. et al., 2003a, ApJS, 148, 1
\bibitem[\protect\citeauthoryear{Bennett et al.}{2003b}]{w1yrfg} Bennett~C.~L. et al., 2003b, ApJS, 148, 97
\bibitem[\protect\citeauthoryear{Bennett et al.}{2011}]{w7yranom} Bennett~C.~L., et al., 2011, ApJS, 192, 17
\bibitem[\protect\citeauthoryear{Bennett et al.}{2013}]{w9yrmaps} Bennett~C.~L., et al., 2013, ApJS, 208, 20
\bibitem[\protect\citeauthoryear{Bernardi et al.}{2005}]{bernardi05} Bernardi~G., Carretti~E., Fabbri~R., Sbarra~C., and Cortiglioni~S., 2005, MNRAS, 364, L5
\bibitem[\protect\citeauthoryear{Bernui}{2009}]{bernui09} Bernui~A., 2009, PRD, 80, 123010
\bibitem[\protect\citeauthoryear{Bernui et al.}{2012}]{bernui06} Bernui~A., Villela~T., Wuensche~C.~A., Leonardi~R., and Ferreira~I., 2006, A\&A, 454, 409

\bibitem[\protect\citeauthoryear{Cabella et al.}{2010}]{cabella10} Cabella~P. et al., 2010, MNRAS, 405, 961
\bibitem[\protect\citeauthoryear{Chingangbam \& Park}{2013}]{chingangbam13} Chingangbam~P., and Park~C., 2013, JCAP, 02, 031
\bibitem[\protect\citeauthoryear{Coles et al.}{2004}]{coles04} Coles~P., Dineen~P., Earl~J. and Wright~D., 2004, MNRAS, 350, 989
\bibitem[\protect\citeauthoryear{Copi et al.}{2007}]{copi07} Copi~C., Huterer~D., Schwarz~D., and Starkman~G., 2007, PRD, 75, 023507
\bibitem[\protect\citeauthoryear{Cruz et al.}{2006}]{cruz06} Cruz~M., Tucci~M., Martinez-Gonzalez~E., and Vielva~P., 2006, MNRAS, 369, 57
\bibitem[\protect\citeauthoryear{Cruz et al.}{2008}]{cruz08} Cruz~M. et al., 2008, MNRAS, 390, 913
\bibitem[\protect\citeauthoryear{Cruz et al.}{2011}]{cruz11} Cruz~M., Vielva~P., Martinez-Gonzalez~E., and Barreiro~R.~B., 2011, MNRAS, 412, 2383

\bibitem[\protect\citeauthoryear{de Oliveira-Costa \& Tegmark}{2006}]{costa06} de Oliveira-Costa A., and Tegmark M., 2006, PRD, 74, 023005
\bibitem[\protect\citeauthoryear{Dineen \& Coles}{2004}]{dineen04} Dineen~P., and Coles~P., 2004, MNRAS, 347, 52

\bibitem[\protect\citeauthoryear{Eriksen et al.}{2004}]{eriksen04} Eriksen~H.~K., Banday~A.~J., Gorski~K.~M., and Lilje~P.~B., 2004, ApJ, 612, 633
\bibitem[\protect\citeauthoryear{Eriksen et al.}{2008}]{eriksen08} Eriksen~H.~K. et al., 2008, ApJ, 676, 10

\bibitem[\protect\citeauthoryear{Finkbeiner}{2003}]{halphamap} Finkbeiner~D.~P., 2003, ApJS, 146, 407
\bibitem[\protect\citeauthoryear{Finkbeiner, Davis \& Schlegel}{1999}]{fdsmap} Finkbeiner~D.~P., Davis~M., and Schlegel~D.~J., 1999, ApJ, 524, 867

\bibitem[\protect\citeauthoryear{Gruppuso \& Burigana}{2009}]{gruppuso09} Gruppuso~A., and Burigana~C., 2009, JCAP, 08, 004

\bibitem[\protect\citeauthoryear{Hansen et al.}{2012}]{hansen12} Hansen~M. et al., 2012, MNRAS, 426, 57
\bibitem[\protect\citeauthoryear{Haslam et al.}{1982}]{haslammap} Haslam~C.~G.~T., Salter~C.~J., Stoffel~H., and Wilson~W.~E., 1982, A\&AS, 47, 1
\bibitem[\protect\citeauthoryear{Hernandez-Monteagudo}{2010}]{monteagudo10} Hernandez-Monteagudo~C., 2010, A\&A, 520, A101
\bibitem[\protect\citeauthoryear{Hinshaw et al.}{2013}]{w9yrcosmo} Hinshaw~G.~F. et al., 2013, ApJS, 208, 19
\bibitem[\protect\citeauthoryear{Howlett et al.}{2012}]{camb2} Howlett~C., Lewis~A., Hall~A. and Challinor~A., 2012, JCAP, 04, 027

\bibitem[\protect\citeauthoryear{Ilic et al.}{2011}]{ilic11} Ilic~S., Douspis~M., Langer~M., Penin~A., and Lagache~G., 2011, MNRAS, 416, 2688
\bibitem[\protect\citeauthoryear{Inoue \& Silk}{2007}]{inoue07} Inoue~K.~I., and Silk~J., 2007, ApJ, 664, 650

\bibitem[\protect\citeauthoryear{Kovacs, Szapudi \& Frei}{2013}]{kovacs13} Kovacs~A., Szapudi~I., and Frei~Z., 2013, Astronomische Nachrichten, 334, 1020

\bibitem[\protect\citeauthoryear{Lacasa et al.}{2012}]{lacasa12} Lacasa~F., Aghanim~N., Kunz~M., and Frommert~M., 2012, MNRAS, 421, 1982
\bibitem[\protect\citeauthoryear{Land \& Slosar}{2007}]{land07} Land~K., and Slosar~A., 2007, PRD, 76, 087301
\bibitem[\protect\citeauthoryear{Leach et al.}{2008}]{leach08} Leach~S.~M. et al., 2008, A\&A, 491, 597
\bibitem[\protect\citeauthoryear{Lewis, Challinor \& Lasenby}{2000}]{camb1} Lewis~A., Challinor~A. and Lasenby~A., 2000, ApJ, 538, 473
\bibitem[\protect\citeauthoryear{Liu, Mertsch \& Sarkar}{2014}]{liu14} Liu~H., Mertsch~P., and Sarkar~S., 2014, ApJ, 789, L29
\bibitem[\protect\citeauthoryear{Lopez-Corredoira, Labini \& Betancort-Rijo}{2010}]{corredoira10} Lopez-Corredoira~M., Labini~F.~S., and Betancort-Rijo~J., 2010, A\&A, 513, A3

\bibitem[\protect\citeauthoryear{Naselsky et al.}{2005}]{naselsky05} Naselsky~P.~D., Chiang~L.-Y., Novikov~I.~D., and Verkhodanov~O.~V., 2005, IJMPD, 14, 1273
\bibitem[\protect\citeauthoryear{Naselsky et al.}{2010}]{naselsky10} Naselsky~P.~D. et al., 2010, Astrophysical Bulletin, 65, 101
\bibitem[\protect\citeauthoryear{Novaes et al.}{2014}]{novaes14} Novaes~C.~P., Bernui~A., Ferreira~I.~S., and Wuensche~C.~A., 2014, JCAP, 01, 018

\bibitem[\protect\citeauthoryear{Planck Collaboration I}{2014}]{planck13} Planck Collaboration I, 2014, A\&A, 571, A1
\bibitem[\protect\citeauthoryear{Planck Collaboration XXIII}{2014}]{plk13anom} Planck Collaboration XXIII, 2014, A\&A, 571, A23
\bibitem[\protect\citeauthoryear{Planck Collaboration IX}{2015}]{plk15cmb} Planck Collaboration IX, 2015, arXiv:1502.05956
\bibitem[\protect\citeauthoryear{Planck Collaboration X}{2015}]{plk15fg} Planck Collaboration X, 2015, arXiv:1502.01588
\bibitem[\protect\citeauthoryear{Planck Collaboration XVI}{2015}]{plk15anom} Planck Collaboration XVI, 2015, arxiv:1506.07135

\bibitem[\protect\citeauthoryear{Rakic, Rasanen \& Schwarz}{2006}]{rakic06} Rakic~A., Rasanen~S., and Schwarz~D.~J., 2006, MNRAS, 369, L27
\bibitem[\protect\citeauthoryear{Rassat et al.}{2007}]{rassat07} Rassat~A., Land~K., Lahav~O., and Abdalla~F.~B., 2007, MNRAS, 377, 1085
\bibitem[\protect\citeauthoryear{Rassat et al.}{2014}]{rassat14} Rassat~A., Starck~J.-L., Paykari~P., Sureau~F., and Bobin~J., 2014, JCAP, 08, 006
\bibitem[\protect\citeauthoryear{Rudnick, Brown \& Williams}{2007}]{rudnick07} Rudnick~L., Brown~S, and Williams~L.~R., 2007, ApJ, 671, 40

\bibitem[\protect\citeauthoryear{Saha}{2011}]{saha11} Saha~R., 2011, ApJ, 739, L56
\bibitem[\protect\citeauthoryear{Sarkar, Datta \& Bharadwaj}{2009}]{sarkar09} Sarkar~T.~G., Datta~K.~K., and Bharadwaj~S., 2009, JCAP, 08, 019
\bibitem[\protect\citeauthoryear{Sawangwit et al.}{2010}]{sawangwit10} Sawangwit~U. et al., 2010, MNRAS, 402, 2228
\bibitem[\protect\citeauthoryear{Short \& Coles}{2010}]{short10} Short~J., and Coles~P., 2010, MNRAS, 401, 2202
\bibitem[\protect\citeauthoryear{Slosar \& Seljak}{2004}]{slosar04} Slosar~A., and Seljak~U., 2004, PRD, 70, 083002
\bibitem[\protect\citeauthoryear{Smith \& Huterer}{2010}]{smith10} Smith~K.~M., and Huterer~D., 2010, MNRAS, 403, 2

\bibitem[\protect\citeauthoryear{Taburet et al.}{2011}]{taburet11} Taburet~N., Hernandez-Monteagudo~C., Aghanim~N., Douspis~M., and Sunyaev~R., 2011, MNRAS, 418, 2207
\bibitem[\protect\citeauthoryear{Tegmark, de Oliveira-Costa \& Hamilton}{2003}]{tegmark03} Tegmark~M., de Oliveira-Costa~A., and Hamilton~A.J., 2003, PRD, 68, 123523
\bibitem[\protect\citeauthoryear{Tomita}{2005}]{tomita05} Tomita~K., 2005, PRD, 72, 103506

\bibitem[\protect\citeauthoryear{Vielva et al.}{2004}]{vielva04} Vielva~P., Martinez-Gonzalez~E., Barreiro~R.~B., Sanz~J.~L., and Cayon~L., 2004, ApJ, 609, 22

\bibitem[\protect\citeauthoryear{Wibig \& Wolfendale}{2015}]{WW15} Wibig~T. and Wolfendale~A.~W., 2015, MNRAS, 448, 1030

\bibitem[\protect\citeauthoryear{Zhang \& Huterer}{2010}]{zhnag10} Zhang~R., and Huterer~D., 2010, Astroparticle Physics, 33, 69
\end{thebibliography}
\end{document}